\documentclass[10pt]{article}

\usepackage{amsmath, amssymb}     % For mathematical symbols and equations
\usepackage{graphicx}             % For including graphics
\usepackage{geometry}             % For setting page margins
\usepackage{fancyhdr}             % For custom headers and footers
\usepackage{titlesec}             % For custom section titles
\usepackage{hyperref}             % For hyperlinks
\usepackage{enumitem}             % For custom lists
\usepackage{setspace}             % For line spacing
\usepackage{authblk}              % For author and affiliation formatting

\titleformat{\section}
  {\normalfont\Large\bfseries}{\thesection.}{1em}{}

\titleformat{\subsection}
  {\normalfont\large\bfseries}{\thesubsection.}{1em}{}

\title{\textbf{Embodied Biocomputing Sequential Circuits with Data Processing and Storage for Neurons-on-a-chip}}

\author[2,3]{Giulio Basso}
\author[1]{Reinhold Scherer}
\author[1]{Michael Taynnan Barros}

\affil[1]{\small{\textit{School of Computer Science and Electronic Engineering, University of Essex, UK}}}
\affil[2]{\small{\textit{Politecnico di Torino, Italy}}}
\affil[3]{\small{\textit{Department of Electrical Engineering, Eindhoven University of Technology, Eindhoven, The Netherlands}}}

\date{}

\begin{document}

\maketitle

\begin{abstract}
With conventional silicon-based computing approaching its physical and efficiency limits, biocomputing emerges as a promising alternative. This approach utilises biomaterials such as DNA and neurons as an interesting alternative to data processing and storage. This study explores the potential of neuronal biocomputing to rival silicon-based systems. We explore neuronal logic gates and sequential circuits that mimic conventional computer architectures. Through mathematical modelling, optimisation, and computer simulation, we demonstrate the operational capabilities of neuronal sequential circuits. These circuits include a neuronal NAND gate, SR Latch flip-flop, and D flip-flop memory units. Our approach involves manipulating neuron communication, synaptic conductance, spike buffers, neuron types, and specific neuronal network topology designs. The experiments demonstrate the practicality of encoding binary information using patterns of neuronal activity and overcoming synchronization difficulties with neuronal buffers and inhibition strategies. Our results confirm the effectiveness and scalability of neuronal logic circuits, showing that they maintain a stable metabolic burden even in complex data storage configurations. Our study not only demonstrates the concept of embodied biocomputing by manipulating neuronal properties for digital signal processing but also establishes the foundation for cutting-edge biocomputing technologies. Our designs open up possibilities for using neurons as energy-efficient computing solutions. These solutions have the potential to become an alternate to silicon-based systems by providing a carbon-neutral, biologically feasible alternative.
\end{abstract}

\section*{Introduction}
Conventional silicon-based computer chips are reaching their limits. Processing speed and circuit complexity are limited by factors such as power consumption, heat dissipation, the degree of miniaturisation, and the manufacturing processes and materials used. \cite{moore2014emerging}. Solutions to further increase viable performance to meet our growing computing needs include research into alternative or new materials, superconductivity, or, more recently, a shift to more sustainable and carbon-neutral approaches such as biological computing (biocomputing).

Deoxyribonucleic acid (DNA), fungi, bacteria, astrocytes and neurons are examples of biomaterials used to deploy biocomputing \cite{grozinger2019pathways}. Which computational biomaterial is most "suitable" depends on the intended application \cite{xue2021biomaterials}. Biomaterial types based on molecules or cells as machines are currently the most viable choices, with the latter being more reliable, faster, and energy efficient \cite{vegh2022role}. For example, 3D cultured biological neurons networks already have demonstrated capabilities to recognise speech \cite{cai2023brain} or to interact with a digital computer system to play computer games \cite{kagan2022vitro,demarse2005adaptive}. Such complex functionality can only be achieved and maintained by adapting the information processing in cells, the cell-cell communications and the structure of the cellular network. Controlling these properties paves the way for human-designed biocomputing, which we call “embodied biocomputing.” This approach implements advanced, controlled digital functions while taking into account the physiological properties and biophysical limits of a living substrate.

While neuronal biocomputing models have the potential to become an alternative to silicon-based computers \cite{barros2021engineering,smirnova2023organoid}, a deeper understanding of how these models can be trained or programmed to perform specific data processing and memory functions is required \cite{bennet2022current}.
While biological systems encode multilevel information, digital data processing is far removed from this characteristic. Moreover, digital technology possesses the capacity to store vast amounts of binary information. To achieve the capabilities of digital storage using biocomputing approaches, one must overcome the challenge of building modular or scalable biocomputing solutions that are also reliable and realisable in-vitro. Biocomputing should draw inspiration from silicon-based digital systems, incorporating Boolean logic and advanced principles of circuit design and synthesis. This approach enables the combination of elementary logic gates into logic circuits capable of performing complex, controllable, and scalable calculations. A corresponding framework is not yet available in biocomputing \cite{basso2022biocomputing}. Moreover, no exploration has been presented into how increasing the computing complexity of biocomputing may lead to an increase in resources within the cellular apparatus, including, for example, its metabolic burden or the relationship between information processing and energy expenditure. By analysing this aspect, researchers would be able to understand the limits of biocomputing, its limitations and its applications. We thus explore, if embodied biocomputing allows us to create neuronal circuits that reliably mimic the behaviour of logic gates, could we then apply the existing logic framework to create biocomputing systems able to perform digital signal storage? And if successful, at what cost do these operations have on biological resource stability?   % and create scalable biocomputing solutions.

Exploratory work has shown that neuronal models can compute outputs using simple logic systems based on AND and OR gates \cite{basso2022biocomputing,knijnenburg2016logic}. Achieving more complex computing circuits beyond cascaded solutions continues to be a major challenge since existing solutions often lack the necessary scalability and efficiency. Current approaches are limited in their ability to integrate multiple logic functions seamlessly. The state-of-the-art neuronal biocomputing circuit includes combinatorial circuits and cascaded circuits of AND and OR gating \cite{adonias2020reconfigurable,adonias2020utilizing}. In this paper, we explore the rational engineering design of neuronal logic systems. Starting with neuronal logic gates, we explore the design of networks composed of these modules to implement synchronous sequential logic, which forms the foundation of silicon-based computer architectures.

We address the challenges posed by simple logic operations and the unscalability and unreliability of neuron-based biocomputing by focusing on the design and engineering of syncronized neuronal networks with specific topologies that implement advanced logic gates (NOT, AND NOT, NAND) in sequential circuits (SR latch, gated SR latch, flip-flop, and D flip-flop). We are the first ones to show that neurons can be used not only for binary data processing but also for data storage by mimicking the same logic circuits used in traditional digital technology to store data. We use in-silico experiments, optimisation and digital circuit principles to design neuronal sequential gates by manipulating its neuron communication networks using principles of embodiment. For our experiments, we used already validated models that account for Spike events, which are the information carrier through the potentiation processes of a neuron membrane, and synapses, which carries information from one neuron to another. We chose Izhikevich neurons with conductance-based excitatory bipartite synapses (i.e only neuron-neuron), instantaneous rise, and single-exponential decay \cite{kim2013coupling} that compose our neuronal units. We also consider different neuron types, including excitatory neurons, i.e., synapses that pass along spikes to another neuron membrane, and inhibitory neurons, which block spikes from being formed based on a synapse event. This is another contribution of our work, since inhibition is now used as a design choice alongside synaptic conductance. Binary information is encoded in periods of spiking ("1") and non-spiking ("0") activities. We rationally designed neuronal synaptic configurations, neuron network topology, and synchronization through neuronal buffers to create digital-like information flow, processing, and storage. We successfully validated the operation of all gates and circuits in-silico. Additionally, we quantified the metabolic states in all our neuron networks to demonstrate the impact of controlled biocomputing on standard energy models, which depend on neuron spiking patterns.
%Our in-silico experiments demonstrate neuronal NAND (NOT-AND) gate by cascading AND and NOT gates, and AND NOT gate-based sequential circuits including a successfully validate a neuronal SR Latch flip-flop digital storage unit. Finally, using interconnected building blocks from our neuronal logic gate library, we implement gated SR Latch (or clocked SR Latch) and D flip-flop memory. We also show the D flip-flop's increased firing frequency by adjusting synaptic weights. 
\textit{The impact of scaling up logic circuits achieves versatility and stable operation with minimal metabolic burden, even when considering data storage circuits.} Through the communication engineering of neurons, biocomputing can be deployed in a versatile manner and possibly lead the way for sustainable computing. Our work forms the basis for powerful biocomputing technologies such as biological finite-state machines and more, giving sustainable computing a beyond-silicon new ally: neurons.

\section*{Results}

% \begin{mdframed}[backgroundcolor=blue!20] 
% {\bf Plan for more results:} \begin{itemize}
%     \item What more delay? increase stimulation duration?
%     \item jitter or phase noise
%     \item logic gate comparison
%     \item biological metric?
%     \subitem Guilio will investigate the manipulation of synaptic strength using drugs
%     \subitem Michael will investigate metabolic stress from the stimulation
%     \item metabolic control. We will think of how we synthetically stop control of neurons to maintain logic gate integrity. 
% \end{itemize}
% \end{mdframed}

Our results validate the operational functioning of each logic structure using their Truth Table, i.e. the correct input-output relationship, to accurately capture their operation. The precise engineering of neuron communication network is highlighted to achieve correctly the desired gate/circuit behaviour. We used the neuron network in-silico experiments, detailed in our Materials and Methods section. We have assumed that the network nodes have pre-established and designed network topologies, which means that no undesired connection must be made to any neuron. This is to ensure that information not only flows as a digital circuit, but to guarantee that communication allows accurate computing outputs.

\subsection*{Neuronal AND NOT-based logic gates} \label{results AND NOT-based}

\begin{figure} [p!]
    \centering
    \includegraphics[width=0.9\textwidth]{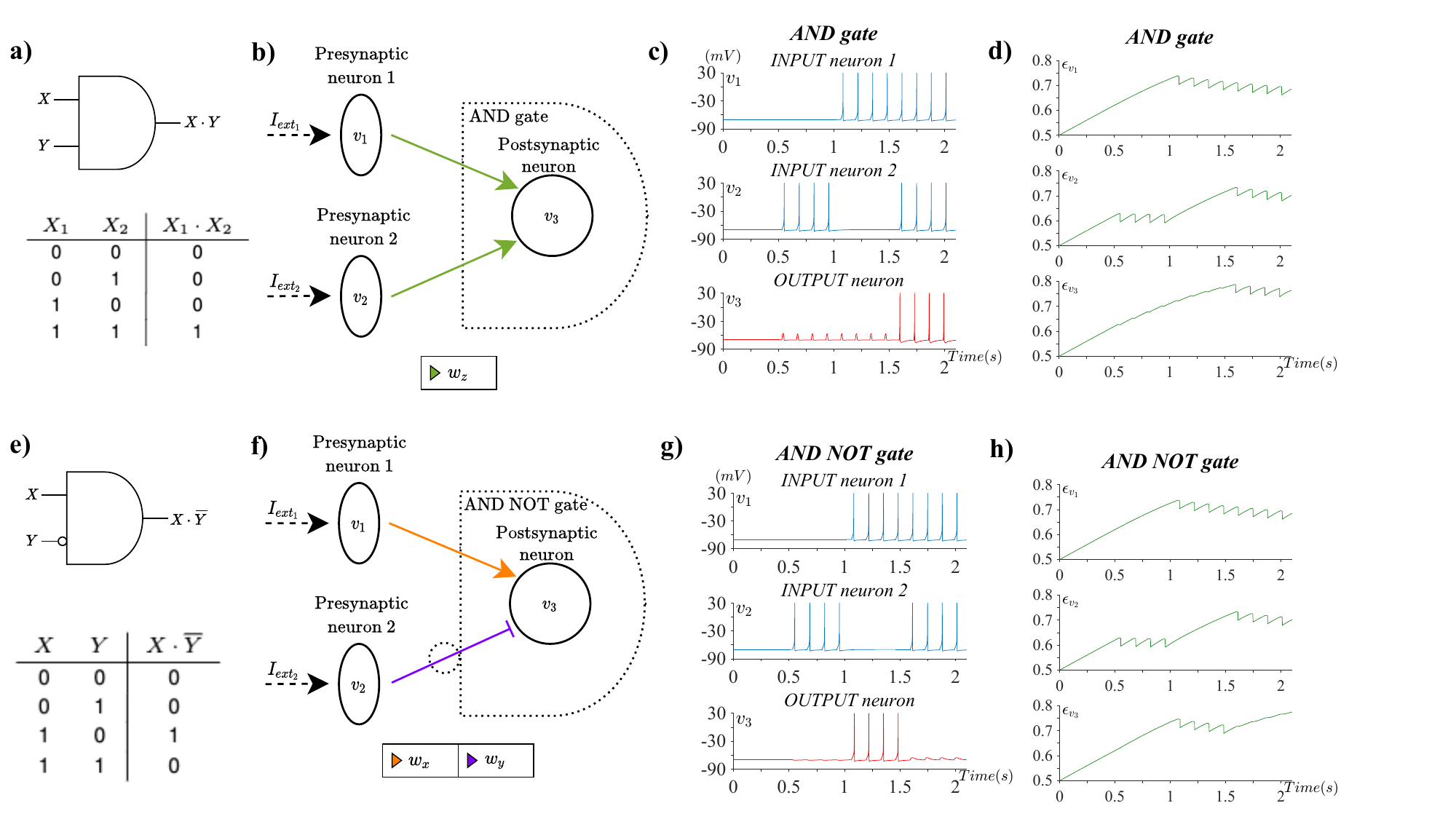}
    \caption{Gating response of the neuronal AND and AND NOT gates. a) The logic gate formal symbolic representation of the AND and below its truth table b) the functional graph of the neurons and connections, including types, composing the AND gate c) the functional validation of the AND gate following the truth table order in a) d) the time-series metabolic burden per neuronal unit e) The logic gate formal symbolic representation of the AND NOT and below its truth table f) the functional graph of the neurons and connections, including types, composing the AND NOT gate g) the functional validation of the AND NOT gate following the truth table order in e) h) the time-series metabolic burden per neuronal unit. For the excitatory synapse, the synaptic parameters are set as: $w_x=0.06$, $\tau_r=19.80$ ms and $\tau_d=26.40$ ms. For the inhibitory synapse, they are chosen as: $w_y=0.18$, $\tau_r=19.80$ ms and $\tau_d=59.40$ ms. The amplitude of the stimulating current is $I=4$ pA. (a-h)}
    \label{result ANDNOT}
\end{figure}

The AND gate is the first neuronal logic gate to be tested, and its operational validation is shown in Fig. \ref{result ANDNOT} (a-d). We fine-tuned the neuronal communication parameters at the synapses (neuron-neuron connection points) so that their synaptic weight, which is the strength of the synapse, is \( w_z = 0.065 \) for both synapses. The spike-raising phase has a time constant \( \tau_r \) equal to 2\% of the inter-spike interval (ISI), which is 2.64 ms, and the decaying phase has a time constant \( \tau_d \) equal to 3\% of the ISI, which is 3.96 ms. The input patterns are chosen according to the traditional incremental order used in truth tables, and the input levels are changed every half of a second. In the first interval, both input neurons are silent, corresponding to input levels [0 0]. In the second and third intervals, only one input neuron is firing, corresponding to [0 1] and [1 0], while in the fourth interval, both input neurons are firing, i.e., [1 1], Fig. \ref{result ANDNOT} (a). Fig. \ref{result ANDNOT} (d) shows the metabolic energy time-series analysis of all neuron units. We consider metabolic burden as the energy expenditure being the ratio of ATP/ADP (more details in our Materials and Methods section). We show that the energy expenditure has a fluctuation behaviour between $0.5$-$0.8$ a.u. cost. The pattern of the fluctuations is visually associated with spike events, since the energy cost is dependent on the neuron membrane voltage. Energy expenditure increases during non-spike events due to a series of possible events, including neurotransmitter packaging and recycling \cite{yuan2018effects}, maintaining ion gradients \cite{engl2015non,zhu2019energy}, cytoskeletal dynamics, and protein synthesis \cite{zhu2019energy}.

The first neuronal logic gate developed is the AND NOT gate, whose results are presented in Fig. \ref{result ANDNOT} (a-d). For this gate, we need to consider both excitatory and inhibitory neuronal types, to account for its more complex Boolean logic. Its excitatory and inhibitory synaptic weights are set as $w_x=0.06$ and $w_y=0.18$ respectively. The synaptic time constant values are chosen as $\tau_r=15\%ISI=19.80$ ms and $\tau_d=20\%ISI=26.40$ ms for the excitatory synapse, and $\tau_r=15\%ISI=19.80$ ms and $\tau_d=45\%ISI=59.40$ ms for the inhibitory synapse. Fig. \ref{result ANDNOT} (h) shows the energy temporal analysis of all neuron units for the AND NOT gate,  also showing fluctuation behaviour between $0.5$-$0.8$ a.u. cost for similar reasons as the AND gate.

\begin{figure} [p!]
    \centering
    \includegraphics[width=0.9\textwidth]{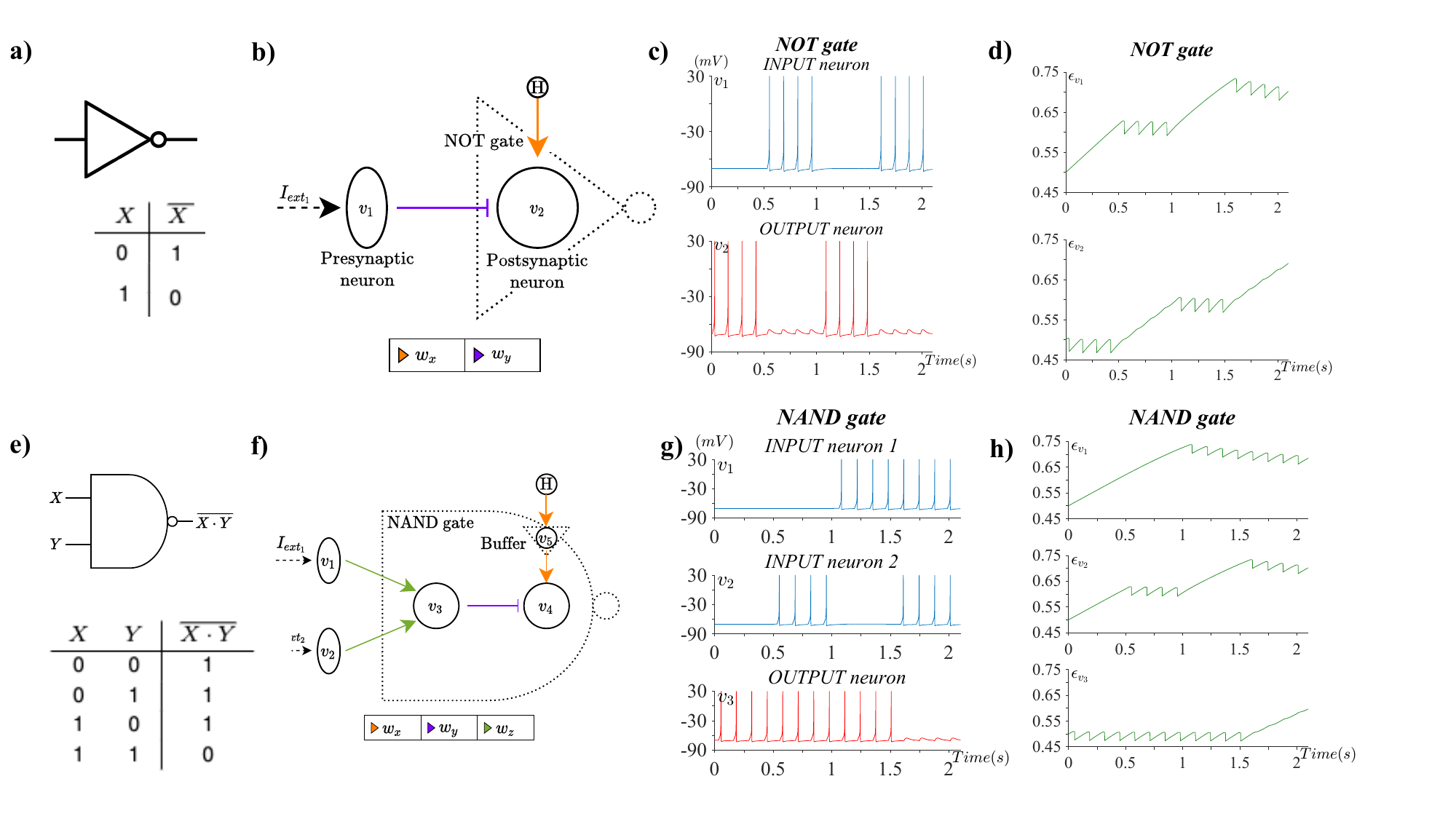}
    \caption{Gating response of the neuronal NOT and NAND gates. a) The logic gate formal symbolic representation of the NOT and below its truth table b) the functional graph of the neurons and connections, including types, composing the NOT gate c) the functional validation of the NOT gate following the truth table order in a) d) the time-series metabolic burden per neuronal unit e) The logic gate formal symbolic representation of the NAND and below its truth table f) the functional graph of the neurons and connections, including types, composing the NAND gate g) the functional validation of the NAND gate following the truth table order in e) h) the time-series metabolic burden per neuronal unit. For the excitatory synapse, the synaptic parameters are set as: $w_x=0.06$, $\tau_r=19.80$ ms and $\tau_d=26.40$ ms. For the inhibitory synapse, they are chosen as: $w_y=0.18$, $\tau_r=19.80$ ms and $\tau_d=59.40$ ms. The amplitude of the stimulating current is $I=4$ pA. Gating response of the neuronal NAND gate. The synaptic parameters of the AND gate are $w_z=0.065$, $\tau_r=2.64$ ms and $\tau_d=3.96$ ms; for the NOT gate excitatory synapse $w_x=0.06$, $\tau_r=19.80$ ms and $\tau_d=26.40$ ms; for the NOT gate inhibitory synapse $w_y=0.18$, $\tau_r=19.80$ ms and $\tau_d=59.40$ ms. The amplitude of the stimulating current is $I=4$ pA.  (a-h)}
    \label{result NAND}
\end{figure}

Afterwards, the neuronal NOT gate is presented, as shown in Fig. \ref{result NAND}. The neuronal NOT gate consists of a special case of the AND NOT gate; hence, the synaptic parameters $w_x$, $w_y$, $\tau_r$, and $\tau_d$ are chosen as in the case of the AND NOT gate. The input pattern is generated as a sequence that toggles between the two states every 0.5 seconds. Here, to account for the ability to switch the level [1] based on the input we add an excitatory input (H node in Fig. \ref{result NAND} b)) that is always at the high level, namely a neuron that is always firing. We thus obtained fluctuations at the outputs at the desired [0] level (Fig. \ref{result NAND} c), which did not lead to spike events, and thus the operation validity of the NOT gate remained correct. In addition, the metabolic energy cost is similar to previous cases.

Lastly, Fig. \ref{result NAND} shows how the neuronal NAND gate changes over time when the above-mentioned incremental input pattern is applied. The AND gate included in the NAND cascaded architecture makes use of the same synaptic parameters of the AND and NOT gates reported before. The NOT gate within the NAND architecture makes use of the same synaptic parameters as the AND NOT gate. Similarly, the neuronal buffers are realised using the synaptic parameters of the excitatory synapse of the AND NOT gate. While the association of the energy cost with spike events still stands, the fluctuations tend to follow the different stability points based on the gate outputs when compared to previously explored AND and NOT gates.

%3 figures in 1 page
%\begin{figure}[htp]% [H] is so declass\'e!
%\centering
%\begin{minipage}[b]{0.48\textwidth}
%\centering
%\includegraphics[width=.75\textwidth]{images/fig chapter 6B/result_ANDNOT.pdf}
%\caption{Example of gating response of the neuronal AND NOT gate. For the excitatory synapse, the synaptic parameters are set as: $w_1=0.06$, $\tau_r=19.80$ ms and $\tau_d=26.40$ ms. For the inhibitory synapse, they are chosen as: $w_2=0.18$, $\tau_r=19.80$ ms and $\tau_d=59.40$ ms. The amplitude of the stimulating current is $I=4$ pA.}
%\end{minipage}\hfill
%\begin{minipage}[b]{0.48\textwidth}
%\centering
%\includegraphics[width=.75\textwidth]{images/fig chapter 6B/result_NOT.pdf}
%\caption{Example of gating response of the neuronal NOT gate. For the excitatory synapse, the synaptic parameters are set as: $w_x=0.06$, $\tau_r=19.80$ ms and $\tau_d=26.40$ ms. For the inhibitory synapse, they are chosen as: $w_y=0.18$, $\tau_r=19.80$ ms and $\tau_d=59.40$ ms. The amplitude of the stimulating current is $I=4$ pA.}
%\end{minipage}\par
%\vskip\floatsep% normal separation between figures
%\includegraphics[width=0.38\textwidth]{images/fig chapter 6B/result_NAND.pdf}
%\caption{Example of gating response of the neuronal NAND gate. The synaptic parameters of the AND gate are $w_z=0.065$, $\tau_r=2.64$ ms and $\tau_d=3.96$ ms; for the NOT gate excitatory synapse $w_x=0.06$, $\tau_r=19.80$ ms and $\tau_d=26.40$ ms; for the NOT gate inhibitory synapse $w_y=0.18$, $\tau_r=19.80$ ms and $\tau_d=59.40$ ms. The amplitude of the stimulating current is $I=4$ pA.}
%\end{figure}

\subsection*{Neuronal latches and D flip-flop} \label{results sequential}

\begin{figure}
    \centering
        \includegraphics[width=\textwidth]{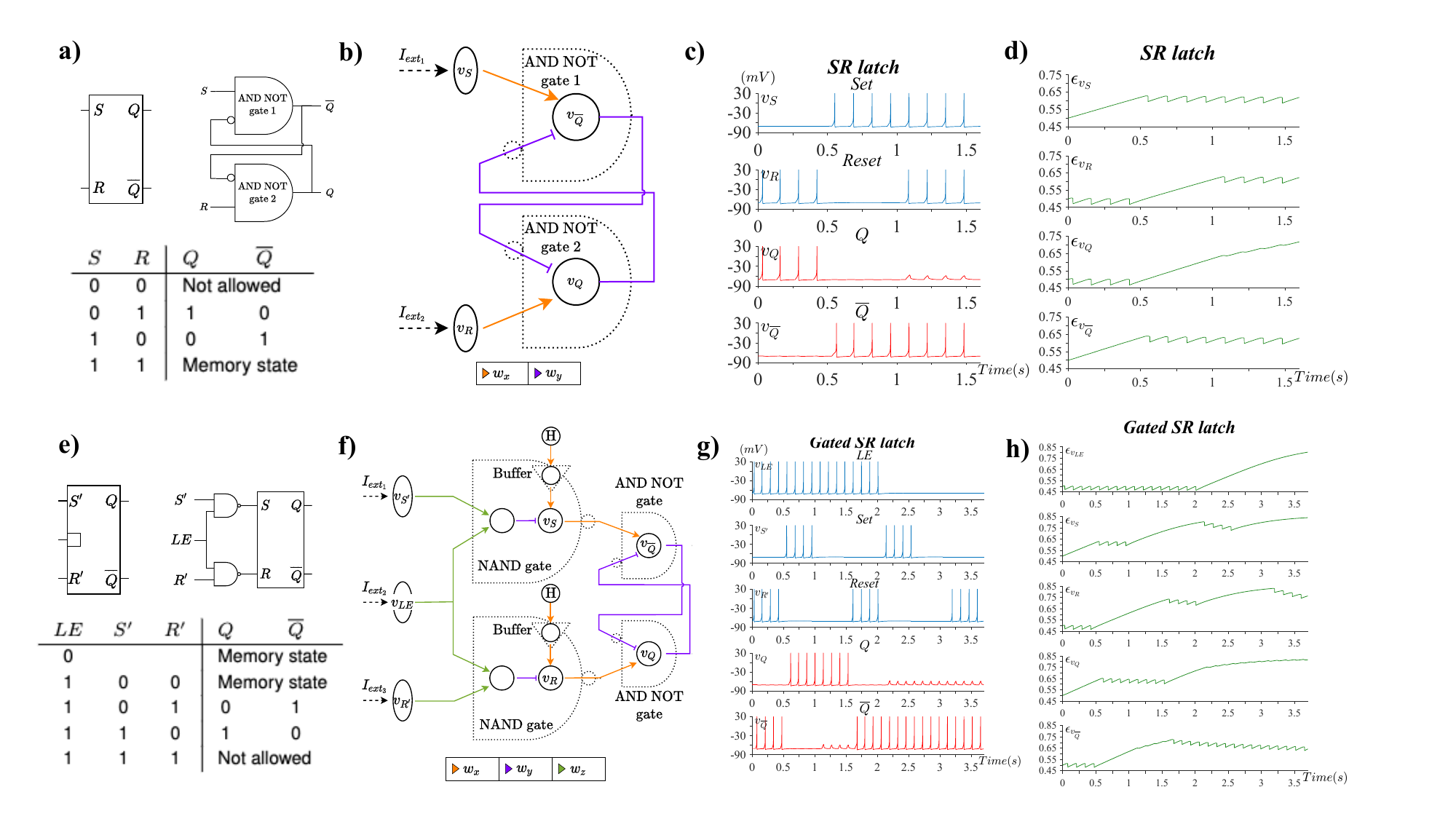} % first figure itself
        \caption{Gating response of the neuronal SR latch and GATED SR latch, with stimulating current $I=4$ pA.  a) The logic gate formal symbolic representation of the SR latch and below its truth table b) the functional graph of the neurons and connections, including types, composing the SR latch c) the functional validation of the SR latch following the truth table order in a) d) the time-series metabolic burden per neuronal unit e) The logic gate formal symbolic representation of the GATED SR latch and below its truth table f) the functional graph of the neurons and connections, including types, composing the GATED SR latch g) the functional validation of the GATED SR latch following the truth table order in e) h) the time-series metabolic burden per neuronal unit.}
        \label{result GATED}
\end{figure}

Fig. \ref{result GATED} shows the dynamics of the neuronal SR latch. Here each membrane potential is labelled with reference to the digital logic signals that it represents. For instance, $v_s$ is the membrane potential that is used as the set signal of the latch. The two outputs of the circuits are represented by the membrane potentials $v_Q$ and $v_{\overline{Q}}$. Note that the condition with both set and reset neurons at the resting state is avoided since it represents the not-allowable condition of the SR latch. The behaviour of the latch is summarised in Fig. \ref{result GATED} a). For instance, lets consider the case with $S=1$ and $R=0$. Since $R=0$ the response of the AND NOT gate 2, namely $Q$, will certainly be 0, because its $X$ input is 0 (see the AND NOT gate Truth Table, Fig. \ref{result ANDNOT} a)). Gate 1 has inputs $S=1$ and $Q=0$, therefore, its output $\overline{Q}$ is 1. Similarly, it can be demonstrated that with $S=0$ and $R=1$, Q is 1, while $\overline{Q}$ is 0. Consequently, this realisation consists of an active low latch, which means that when it is activated by setting the set equal to 1 and reset to 0, its output $Q$ is at the low level. Now considering the case with $S=1$ and $R=1$, since each of the two gates has $X$ input as 1, they behave as a NOT gate. Suppose that in the instant before the setting of $S=1$ $R=1$ the latch presented outputs $Q_{-1}$=1 and $\overline{Q}_{-1}$=0. Once chosen S and R to 1, gate 1 has inputs $S=1$ and $Q_{-1}=1$, and so $\overline{Q}$ is 0; gate 2 has inputs $R=1$ and $\overline{Q}_{-1}$=0 and so $Q$ is 1. Therefore, the outputs are unchanged. Similarly, one could demonstrate the case with $Q_{-1}$=0 and $\overline{Q}_{-1}$=1. This inputs configuration is called the \emph{memory state}, because the latch stores the 1-bit information which represents its state. Finally, in the case with $S=0$ and $R=0$, since both $X$ inputs are 0, both $Q$ and $\overline{Q}_{-1}$ are 0. This combination is often called \emph{not allowed} condition (or forbidden condition), because now the fundamental relationship that $\overline{Q}$ corresponds to the negation of $Q$ is not valid anymore. Even though this condition does not damage the device, it must be avoided because the latch is not following the desired logic behaviour.

Next, the results related to the neuronal gated SR latch are illustrated in Fig. \ref{result GATED} (e-h). The membrane potential $v_{LE}$ is associated to the $LE$ signal which drives the front gating stage. Differently from the previous SR latch, now the not allowed condition corresponds to the case with both set and reset neurons at the high level, and hence it is excluded in the input pattern. Observing the NAND Truth Table \ref{result GATED} e), it can be noticed that if $LE=0$ the NANDs outputs are always 1. Hence the following SR latch is forced in the memory state, namely it is not transparent respect the inputs. On the contrary if $LE=1$, the NANDs outputs correspond in the negation of their other inputs ($S'$ or $R'$). Therefore in that condition, the SR latch receives the inputs $\overline{S'}$ and $\overline{R'}$, and so its behaves as an active high latch, i.e. the setting of $S'=1$ and $R'=0$ results in a high output $Q$.

\begin{figure}
\centering
    \includegraphics[width=0.9\textwidth]{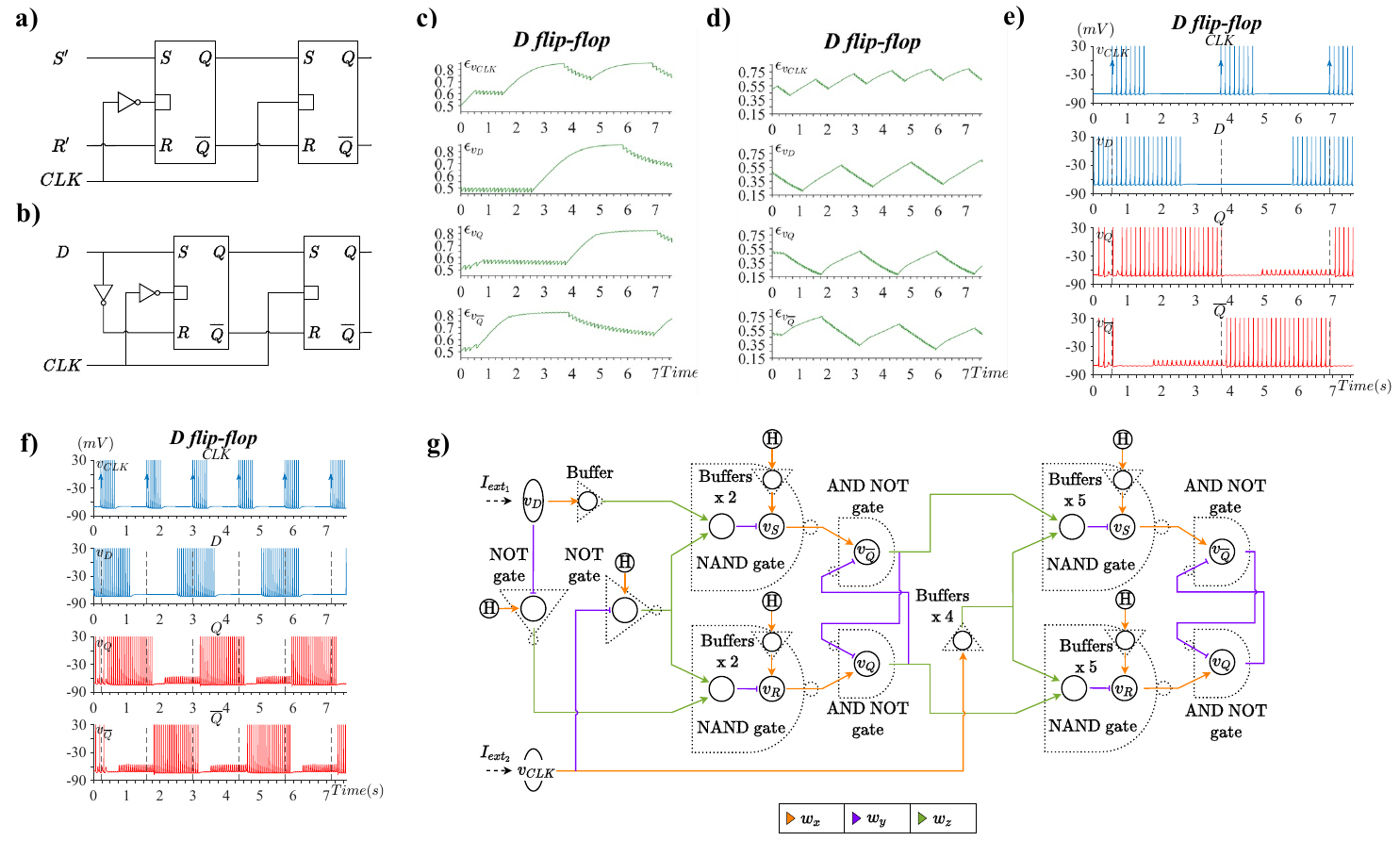}
    \caption{Gating response of the neuronal flip-flop and D flip-flop with stimulating current $I=4$ pA and $I=7$ pA. a) symbolic representation of the flip-flop b) the symbolic representation of the D flip-flop c) the time series burden of the D flip-flop inputs and output neuronal units with $I=4$ pA d) the time series burden of the D flip-flop inputs and output neuronal units with $I=7$ pA e) the functional validation of the D flip-flop inputs D and CLK and outputs Q and $\overline{Q}$ with $I=4$ pA  f) the functional validation of the D flip-flop inputs D and CLK and outputs Q and $\overline{Q}$ with $I=7$ pA g) the functional graph of the neurons and connections, including types, composing the D flip-flop. }
    \label{result D FLIPFLOP7}
\end{figure}

Most digital flip-flops are obtained from a fundamental architecture, called \emph{primary-secondary}. Fig. \ref{result D FLIPFLOP7} shows a primary-secondary flip-flop, which comprises two gated SR latches connected in cascade. The primary transmits the input values, and the secondary samples the outputs of the primary or holds them in memory. This is achieved with a shared clock signal $CLK$, which is directly sent to the secondary and inverted to the primary. When the clock signal is low ($CLK=0$), the $LE$ signal seen by the primary latch is high, and so changes in the input levels influence the primary outputs. On the contrary, the secondary is forced into the memory condition because its $LE$ is low, hence its outputs remain stable. When the clock signal switches to the high level ($CLK=1$), the primary is locked and maintains the previous output levels. At the same time, the secondary becomes transparent, and sets its outputs according to the received inputs, which are the fixed outputs held by the primary. Since after the rising of the clock, the primary is not sensitive to the input variations anymore, the flip-flop outputs correspond to the input values at the instant of the rising edge. Therefore, this device represents a positive-edge flip-flop because it is sensitive at the rising edge of the clock. Similarly, it could be demonstrated that by placing the not gate only in the secondary clock, the device behaves as a negative-edge flip-flop. Although the clock signal regulates the operation of the flip-flop, the device can still encounter an invalid state when both inputs are high. Fig. \ref{result D FLIPFLOP7} displays the membrane potentials of the neuronal D flip-flop. In that case, the input pattern consists of the single D signal, whose role is played by membrane potential $v_D$. As an example of a synchronous edge-sensitive circuit, this device is triggered by a clock signal, depicted by the membrane potential $v_{CLK}$. Specifically, this implementation corresponds to a positive-edge flip-flop. Hence, the triggering rising edges of the clock are highlighted with arrows in $v_{CLK}$, and reported with dashed lines in the other signals. The D flip-flop is also tested with an increased stimulating current equal to $I=7$ pA, as shown in Fig. \ref{result D FLIPFLOP7}. The values of the synaptic weights are recomputed when $I=7$ pA following the same procedures explained for AND and AND NOT gates. Thus, the obtained weights of the AND NOT gates are $w_x=0.11$ and  $w_y=0.67$ for the excitatory and inhibitory synapses, respectively, while the weights of the AND gates are set at $w_z=0.099$. For the SR latch, the gated SR latch, and the flip flop, energy costs, while also varying between $0.5$-$0.8$ may present increasingly complex fluctuations. The fluctuations not only seem to be dependent on the voltage but also on the network topology complexity that is observed in each of the circuits.

%\begin{figure}%[t!]
%    \centering
%    \includegraphics[width=0.7\textwidth]{images/fig chapter 6B/result_SR.pdf}
%    \caption{Example of gating response of the neuronal SR latch, with stimulating current $I=4$ pA.}
%    \label{result SR}
%\end{figure}

%\begin{figure}%[t!]
%    \centering
%    \includegraphics[width=0.8\textwidth]{images/fig chapter 6B/result_GATED.pdf}
%    \caption{Example of gating response of the neuronal gated SR latch, with stimulating current $I=4$ pA.}
%    \label{result GATED}
%\end{figure}

\subsection*{Discussion}

The use of inhibition provides an effective tool for more complex circuits since our library of gating responses has been expanded with the realisation of AND NOT, NOT and NAND gates.  The realisation of each gate involved different choices regarding the synaptic model parameters. For the AND NOT gate, we can observe that the value of the weights related to the inhibitory synapse ($w_y=0.18$) is higher than that of the excitatory synapse ($w_x=0.06$).
The reason for this could be related to the fact that when excitatory and inhibitory stimuli occur nearby, the inhibition must overcome the excitation, resulting in the resting state.
Specifically for the AND NOT gate, the decay time constant of the inhibitory synapse ($\tau_d=45\%ISI=59.40$ ms) is chosen with greater respect the one of the excitatory synapse ($\tau_d=20\%ISI=26.40$ ms), because we addressed to the issues related to the mismatched synchronization by relaxing the dynamic of the inhibition. Furthermore, it can be noticed that the synaptic time constants related to the AND gate ($\tau_r=2\%ISI=2.64$ ms and $\tau_d=3\%ISI=3.96$ ms) are significantly smaller than in the case of the AND NOT gate. Nonetheless, the chosen values of synaptic time constants belong to the range of physiological values found in literature. For instance, the fastest AMPA receptors display a time constant of 0.18 ms, while the NMDA receptor of pyramidal cells is up to 89 ms \cite{roth2009modeling}. Overall, Izhkevich neurons are well-known for accurately capturing the dynamics of neuron membrane potential and the recovery variable, which represents the activation of ion channels in the cell membrane \cite{almog2016realistic}. Specifically, the Izhikevich model is able to exhibit 20 of the most prominent features of biological neurons, including tonic spiking, intrinsic bursting, chattering, and more \cite{de2023analysis}. This is in contrast to the LIF model, which is a much simpler model that can only capture basic integrate-and-fire behavior. In our simulations, we have operated with tonic spiking range, which is present in midbrain dopaminergic neurons \cite{radulescu2010mechanisms}, thalamic relay neurons \cite{connelly2017variable}, and striatal spiny projection neurons \cite{matityahu2022tonic}. The choice of different spiking patterns and different neuron types can further enhance our biocomputing designs in an open-ended manner, and more investigation is needed to elicit the mapping of computing building blocks to neuronal circuit set ups.

The versatility of our designed sequential logic circuits is represented by the fact that larger circuits are simply built starting from the elementary logic gates, which are already trained. Therefore, synaptic parameters can be transferred in a modular fashion. This leads to scaling opportunities beyond flip-flops, as modules can be interconnected to create increasingly larger and more complex logic circuits. However, there is a caveat to consider: the number of design variables that need to be incorporated, including latency profile, buffers, and clock configuration. Variations in excitatory and inhibitory inputs across different layers, especially in complex neuron network structures, cause an increase in spike latency, leading to synchronisation issues within the circuit. All our designs are sensitive to such phenomena, leading to logic errors or even operational failures when desynchronization happens. That is why we considered buffers in most of our designs. With the exception of AND and AND NOT gates, buffers were added to excitatory neurons to synchronise misaligned excitatory and inhibitory output spikes. Moving forward towards scalable circuits, we need a solution to the challenge of neuronal network properties with different neuron types, latency quantification, and formal buffering guidelines, which opens the door to exciting novel methods of biocomputing scaling mechanisms, possibly incorporating embodiment as a design methodology. In addition, to capture the essence of clock-based solutions, our D-flip-flop solution includes a clock input, which is yet to be fully validated experimentally. We however, are confident that such a concept can be transferred from clock-regulation circuits already observed in-vivo \cite{hastings1998brain}. Clocks are essential for digital computing solutions, and we predict that they will also determine the success of biocomputing solutions, since the latter also needs processes that dictate information flow, which can be achieved through the implementation of clocks. 

%The reason of this is related to the fact that the neuronal flip-flop is not initialized yet, and hence the output is well-defined only after the first clock edge. Similarly in digital electronics, when the flip-flop is turned on, it can not be predicted if the initial state is the low or high level. This is the reason why often supplementary inputs, often called preset and clear are used to initialize the device. In Fig. \ref{result D FLIPFLOP7} we can observe that, as in digital flip-flops, the output of the neuronal flip-flop changes after a certain delay respect the clock edge. This is defined as the propagation delay, which is the time needed between the clock edge and the resulting change at the flip-flop output $Q$ \cite{tessier2003understanding}. Finally Fig. \ref{result D FLIPFLOP7} reports and example of the behaviour of the D flip-flop with increased stimulating current $I=7$ pA, which results in an higher neurons firing frequency. Furthermore, other open questions result from the fact that the ultimate networks topology of neuronal populations is often unknown and it strictly depend on the neuronal activity due to the processes of neuronal plasticity.

Experimental biocomputing has achieved remarkable advancements through the years, however, based on the results of this paper, experimental biocomputing need to show how scalable and reliable they are, not only for neurons but for all other types of biocomputing systems. Current approaches have several drawbacks, including a lack of scalable designs, a lack of packaged solutions, and an over-reliance on complex predictive tools to manage high-intensity information flow and operational accuracy \cite{grozinger2019pathways}. We envision that a neuron-based biocomputing experimental framework will be focused on neurons-on-a-chip. This technology offers the ability to isolate neurons, control synaptic connections, and interface with stimulation and recording tools. Neurons-on-a-chip topology control mechanisms can optimise in-chip biocomputing and further provide information flow control. We are the first ones to show that sequential logic circuits are possible in a versatile fashion, expanding what has been achieved in biocomputing and demonstrating how digital information can be processed and stored. Using sequential logic circuits as building blocks, cascading modules can create larger computing systems. We predict that neurons-on-a-chip will lead to experimental validation of our biocomputing solutions. The ability to also incorporate storage can take biocomputing using neurons from mere function towards much larger, complex, scalable and modular computing. When this is fully achieved, experimental efforts will lead to a world-wide revolution in using biology as viable computing solutions, which have even already been proposed for controlled robots \cite{novellino2007connecting}, analysing new drugs \cite{smirnova2023organoid}, or even playing ping-pong \cite{kagan2022vitro}. The versatile and groundbreaking designs presented in this paper herald an exciting and illuminating journey into scalable and realisable neuron-based embodied biocomputing sequential circuits. This paves the way for integrating human-designed intelligence into living biological machines, opening up a future where the fusion of technology and biology redefines the boundaries of computing.
%The first neuronal sequential logic circuit is a network of neurons in which input signals are generated by a first layer, elaborated according to a network of synaptic connections by intermediate layers, and mapped into outputs by an output layer.

\section*{Materials and Methods}

\subsection*{In-silico experiments}
In-silico experiments depicted in our Results section was computed using MATLAB R2021b Simulink. The equations are discretized using the explicit Euler method with step size $dt=0.5$ ms. The Simulink parameter sample time, which indicates when a Simulink block produces outputs and updates its internal state, is chosen to be equal to $dt$. The Izhikevich neurons involved in the model are characterised by tonic spiking behaviour. The reversal potential $E_{syn}$ is set as $0$ for excitatory synapses and as $-75$ for inhibitory synapses. All the logic circuits are driven by an external stimulating current equal to $I=4$ pA. Finally, an example of logic circuit with higher stimulating current $I=7$ pA is also provided. 

\subsection*{Engineering Neuron Communications} \label{section engneucoms}
One of the key aspects of the design of neuronal logic gates is the possibility of cascading, that is, connecting more of them using the gates output as the inputs of another consecutive gate. Hence, here we formalise the neuronal logic gates as being made only by the postsynaptic neuron and its synapses with the adjusted weights $w_i$. The inputs of the logic gates are represented by the membrane potentials $v_1$ and $v_2$, while the external current $I_{ext_i}$ are only used in the first layer to communicate the input pattern to the network. As $v_3$ is their output.

\subsection*{Neuronal AND NOT gate and NOT gate} \label{section AND NOT and NOT}

The AND gate is the logic gate that implements the logical conjunction. That means that its output is at the high only if both of its inputs are at the high level, as indicated in the Truth Table in Fig \ref{result ANDNOT} a). Similarly, a neuron that fires only if it concurrently receives two presynaptic stimuli implements the same boolean rule, and we define it as neuronal AND gate. Fig. \ref{result ANDNOT} b) describes the functional scheme of the neuronal AND gate, %It consists in two presynaptic neurons, with their membrane potentials $v_1$ and $v_2$ as the logic inputs, and a postsynaptic neuron, whose membrane potential $v_3$ is the logic output. 
which consists in 3 neurons. Two presynaptic neurons are connected to the same postsynaptic neuron through two distinct excitatory synapses. The dashed arrows indicate the external stimulating current $I_{ext_i}$, always chosen as step or rectangular functions. 
%Neurons which cover only a presynaptic role (i.e. they do not form any synapses in which they represent the postsynaptic neuron) are depicted by ellipses, characterised by their membrane potential variable. The circles stand for the postsynaptic neurons. The equivalent traditional symbol of the logic gate is reported with dotted line. 
For neuronal AND gate design, we need to regulate the inputs' influence on the synapses so that the firing of just one presynaptic neuron is not enough to trigger the output response. We propose the fine regulation of the synaptic weights $w_i$, i.e. the strength of the synapses, namely how much a presynaptic input affects its corresponding postsynaptic neuron. Since the AND gate acts symmetrically the two synaptic weights $w_i$ need to be equal. The value of $w_i$ is here computed empirically between real values of $0$ and $1$, searching for the minimum value of $w_i$ that makes the output neuron firing in the condition with both inputs high. We refer to the weight value of the AND gate's synapses obtained with such empirical procedure as $w_z$.
 
In \cite{yoder2009explicit} L. Yoder theorized a novel type of logic gates, particularly suitable for biocomputed implementations with neurons, that can be flexibly used to develop several computational solutions. As shown in Fig. \ref{result ANDNOT}, it consists in an AND gate in which one of its two inputs is inverted. L. Yoder referred to this special type of logic gate as \emph{AND NOT gate}. Considering a traditional AND gate, its output is 1 if and only if both inputs are 1. Instead, for the statement $X$ AND NOT $Y$, the high output requires $X$ to be 1, while $Y$ needs to be 0, because of the logical negation. In any other cases the output is 0, as depicted in the Truth table (Fig. \ref{result ANDNOT} a). Notice that, unlike traditional gates, as AND and OR gates, the AND NOT gate is not symmetrical in respect its inputs. The gating behaviour of the AND NOT gate could be considered similar to a neuron response considering a neuron with one excitatory synapse and one inhibitory synapse, along the following: i) counterbalance of excitation elicited from excitatory stimuli from an inhibitory stimuli ii) rebound spiking (\cite{izhikevich2004model}) are not observed and iii) a single presynaptic input can generate postsynaptic response alone.
% \begin{itemize}
%     \item the inhibitory effect of inhibitory stimuli is able to perfectly counterbalance the excitation elicited by excitatory stimuli. That means that if the neuron simultaneously receives both an excitatory input and an inhibitory input, the output response is repressed, and so the neuron remains at the resting state;
%     \item the inhibitory input is not able to make the postsynaptic neuron firing, i.e. particular case of spiking behaviour, as rebound spike \cite{izhikevich2004model}, are excluded;
%     \item the firing of the only presynaptic excitatory neuron is enough to elicit the firing of the postsynaptic neuron. This can be achieved with strong synaptic strength.
% \end{itemize}
%Therefore, the neuron will fire if and only if the excitatory input is firing and the inhibitory input is quiet, while in any other case, it will remain in the resting state. This neuronal response is curiously similar to the truth table of the AND NOT gate. 
%The excitatory input can be considered as the logic variable $X$, while the inhibitory input as the logic variable $Y$. 
Hence, a neuronal AND NOT gate can be developed considering a neuron with an excitatory synapse and an inhibitory synapse, as illustrated in Fig. \ref{result ANDNOT}, where the bar-headed arrow stands for the inhibitory connection. From now on, we will refer to the weight value of the excitatory AND NOT gate synapse obtained from an empirical estimation procedure and referenced as $w_x$. The synaptic weight associated to the inhibitory synapse must be set such that if both excitatory and inhibitory inputs are high, then the postsynaptic neuron must be in a resting state. Now, the value of the inhibitory weight $w_y$ can be derived from $w_x$ previously found, setting both inputs high and increasing the inhibitory weight until the postsynaptic response is repressed.
% \begin{figure} [t!]
%     \centering
%     \includegraphics[width=0.6\textwidth]{neuronal AND not.pdf}
%     \caption{Functional scheme of the neuronal AND NOT gate. The excitatory synapse is characterised by the synaptic weight $w_x$, while the inhibitory synapse by $w_y$.}
%     \label{AND NOT scheme}
% \end{figure}

The NOT gate consists of a single input gate, which implements the logical negation. That means that if its input is 0, the output response is 1, and vice versa. It can be noticed that the NOT gate can be directly obtained from the neuronal AND NOT gate \cite{yoder2020neural}. Indeed, looking at the AND NOT gate Truth Table \ref{result ANDNOT}, the output of the last two conditions, where $X=1$, corresponds to the negation of $\overline{X}$. Therefore, an AND NOT gate in which the input $X$ is always high implements a NOT gate that respects the input $\overline{X}$.
Thus, a neuronal NOT gate can be designed by using an excitatory input that is always at a high level. Spontaneously and continuously active neurons are demonstrated to exist in the brain \cite{yoder2009explicit}\cite{mccormick2005neuronal}. For instance, they are involved in the awake condition, and sleep requires their inhibition. Notice that, since the logic NOT gate has only one input, here the "effective" input is represented by the inhibitory input. Fig. \ref{result NAND} depicts the functional scheme of the NOT gate. 
%The continuously active neuron is implemented as a presynaptic neuron whose external stimulating current is always ON. Since we are interested only in the state of the inhibitory input, the continuously firing neuron is simply represented by an 'H', which stands for high level. 
Since the NOT gate shares the same network structure of the AND NOT gate, the synaptic weights assume the same values $w_x$ and $w_y$.

% \begin{figure} [t!]
%     \centering
%     \includegraphics[width=0.6\textwidth]{NOT gate.pdf}
%     \caption{Functional scheme of the neuronal NOT gate.}
%     \label{NOT scheme}
% \end{figure}

\subsection*{Network synchronisation and neuronal buffers} \label{syncNAND}
Our preliminary tests displayed that inhibition is highly sensitive to input timing. Indeed, considering the neuronal AND NOT gate, if the excitatory and inhibitory inputs are not synchronized the inhibition is not effective. %Therefore, even though both inputs are high, the output response could not be the desired resting state, but rather the firing state. 
Synchronisation becomes a critical issue in the case of logic gates cascading or circuits. Considering more logic gates consecutively connected, the implementation of each Izhikevich neurons elaborates the received synaptic inputs to compute the resulting membrane potential. Therefore, each neuron block inevitably introduces a delay on the chain of signals transmission. As a consequence, input synchronisation is not always ensured. Here we address the aforementioned challenges by i) the time constants related to the synapses are increased, such that the variations of the conductances require more time and so the inhibitory and excitatory effects last longer ii) using an approach inspired by digital electronics, we introduce the use of \emph{neuronal buffers}. In digital electronics, a digital buffer is a gate whose output state is equal to its input state with impedance matching and for restoring degraded digital signals after propagation over long distances. For our purpose, the buffer can be used to simply introduce a delay in the input signals and counterbalance their asynchronization. A neuronal buffer can be developed as a neuron with a unique excitatory synapse, with a value of synaptic weight chosen such that if its associated input neuron is firing, also the output neuron fires. 
%Notice that the AND NOT gate also receives only one excitatory input that is able to make its output fire. 
Hence, the buffer synaptic weight can be chosen as the weight of the excitatory synapse of the AND NOT gate $w_x$. The input synchronisation along the neuronal circuits is guaranteed, ensuring the following condition: when more gates are connected together, logic gate inputs must pass through the same number of neuron layers in order to introduce the same delay. Whenever a branch of the neuronal circuit does not respect this condition, neuronal buffers are added. 
%Logic circuits with poor synchronisation can be fixed with the use of neuronal buffers.

% \begin{figure}[b]
% %\begin{center}
% \begin{tabularx}{\textwidth}{*{2}{>{\centering\arraybackslash}X}}
%    \centering
% \includegraphics[width=0.35\textwidth,valign=m]{NAND symbol.pdf} % <-- valing is chanded from T to m
% &
%        \begin{tabular}{c  c | c}
%         $X$ & $Y$ & $\overline{X\cdot Y}$ \\
%         \hline
%         0 & 0 & 1 \\
%         0 & 1 & 1 \\
%         1 & 0 & 1 \\
%         1 & 1 & 0 
%         \end{tabular}          \\ % <-- added new row for captions
% \captionof{figure}{Symbol of the NAND gate}
% &     
%     \captionof{table}{\label{table NAND}Truth Table of the NAND gate. The NAND operation can be written as $\overline{X\cdot Y}$.}
% \end{tabularx}
% % \end{center}
% \vspace{-10mm}
%  \end{figure}
 
\subsection*{Neuronal Cascaded Circuits} \label{neuronalCascades}
%  \begin{figure}
% \centering
%   \begin{subfigure}{0.5\textwidth}
%     \includegraphics[width=\textwidth]{NAND1.pdf}
%     \caption{}
%     \label{NAND1}
%   \end{subfigure}
%   \hfill %%
%   \begin{subfigure}{0.5\textwidth}
%     \includegraphics[width=\textwidth]{NAND2.pdf}
%     \caption{}
%     \label{NAND2}
%   \end{subfigure}
%   \caption{Functional scheme of the neuronal NAND gate. The NAND gate can be obtained using an AND and NOT gates cascading, as in Fig. \ref{NAND1}. The synchronization of the inputs of such neuronal circuit can be achieved with the addition of a neuronal buffer, as shown in Fig. \ref{NAND2}.}
%   \label{}
% \end{figure}
Digital electronics often benefits from the use of NAND gates. Specifically, their importance is related to the fact that any Boolean function can be implemented with a combination of NAND gates. This powerful feature, which also the NOR gates, is called functional completeness. The NAND gate is defined as an AND gate followed by a negation operation. Consequently, its output is low only if both its inputs are high, while it is high in any other cases (Fig. \ref{result NAND} e)). THe straightforward implementation of the neuronal NAND gate can be achieved by cascading a neuronal AND gate with a NOT gate. This simple cascaded structure is depicted in Fig. \ref{result NAND} f). The same values of synaptic weights $w_x$ and $w_y$ of the AND NOT gate implementation can be used. Instead, we refer to the synaptic weight value of the AND synapses as $w_z$. On the synchronisation of the scheme in Fig. \ref{result NAND} f), consider the neuron related to the NOT gate, with membrane potential $v_4$. Its inhibitory synaptic input passes through two layers of neurons, specifically the one associated to the presynaptic neurons (with membrane potentials $v_1$ and $v_2$) and the one related to the AND gate (with membrane potential $v_3$). On the contrary, the excitatory synaptic input passes only through one layer of neurons, which is the continuously firing neuron. Hence, the synchronisation condition is not fulfilled, and the logic circuit could not follow the desired behaviour. The circuit can be adjusted by adding a neuronal buffer in the branch related to the excitatory synapse of the NOT gate, as depicted in Fig. \ref{result NAND} f). 
%Since now both synaptic inputs pass through two layers of neurons, the condition is satisfied.

%SECTION
\subsection*{Neuronal latches}
All the neuronal implementations of logic circuits considered so far make use of combinatorial logic. In combinatorial circuits, their outputs are fully defined by the present values of the inputs. Therefore they could be defined as static, meaning that they do not depend on the previous values of inputs and outputs. Differently in sequential logic circuits, the definition of the outputs relies also on the sequence of past inputs. For this reason the logical outputs are often referred as \emph{states}, which recall the storage capability of sequential logic.
% \begin{figure} [b]
%     \centering
%     \includegraphics[width=0.35\textwidth]{SR logic scheme.pdf}
%     \caption{Logic circuit of the SR latch based on AND NOT gates.}
%     \label{SR logic scheme}
% \end{figure}

% \begin{figure}
% %\begin{center}
% \begin{tabularx}{\textwidth}{*{2}{>{\centering\arraybackslash}X}}
%    \centering
% \includegraphics[width=0.3\linewidth,valign=m]{SR symbol.pdf} % <-- valing is chanded from T to m
% &
%       \begin{tabular}{c  c | c  c}
%         $S$ & $R$ & $Q$ & $\overline{Q}$ \\
%         \hline
%         0 & 0 & \multicolumn{2}{l}{Not allowed} \\
%         0 & 1 & 1 & 0 \\
%         1 & 0 & 0 & 1 \\
%         1 & 1 & \multicolumn{2}{l}{Memory state} 
%         \end{tabular}          \\ % <-- added new row for captions
% \captionof{figure}{Symbol of SR the latch.}
% &     
%     \captionof{table}{\label{table SR}Truth Table of the SR latch.}
% \end{tabularx}
% % \end{center}
% \end{figure}

One of the simplest sequential circuits is the \emph{set-reset latch} (SR latch). It consists in a 1-bit memory, in which its state is defined asynchronously. SR latches are commonly built using NOR gates or NAND gates. Even though we already presented a model of neuronal NAND, here we discuss the realisation of SR latch based on AND NOT gates \cite{yoder2020neural}. NAND gate implementation involves a larger number of neurons that respect the AND NOT gate, and hence its computational cost is higher. The logic scheme of the AND NOT gate-based SR latch is shown in Fig. \ref{result GATED}. The SR latch comprises two AND NOT gates, in which each inverted input (the $Y$ input of the original AND NOT gate) is obtained using the output of the other gate. Here, the not inverted inputs (the $X$ input) are called set ($S$) and reset ($R$). The two neuronal AND NOT gates have mutual inhibitory feedback, meaning that the response of the output neuron of each gate is used as the inhibitory input of the other gate. The functional scheme of the neuronal SR latch is displayed in Fig. \ref{result GATED} b). The synaptic weights of the excitatory and inhibitory synapses can be chosen according to the AND NOT gate implementation. A critical point of the neuronal latch is that here, since inhibitory inputs are obtained from the outputs of the gates, they will always be delayed relative to the excitatory inputs. Indeed, the rule based on the number of neuronal layers is not respected, and there is no neuronal buffer configuration that can re-equalize them. As instance, if we add a buffer in the branch related to the excitatory input of AND NOT gate 1, this gate will now satisfy the rule, but the inhibitory input of AND NOT gate 2 will be delayed even more. The strategy that we adopt here to solve this issue consists in making the inhibitory effect last longer and relaxing the synchronisation requirement. This can be achieved exploiting the temporal summation mechanism, obtained with increased time constants of the inhibitory synapses.

% \begin{figure}[t]
%     \centering
%     \includegraphics[width=0.45\textwidth]{SR latch.pdf}
%     \caption{Functional scheme of the neuronal SR latch.}
%     \label{SR latch scheme}
% \end{figure}

% \begin{figure}[b]
%     \centering
%     \includegraphics[width=0.3\textwidth]{Gated SR logic scheme.pdf}
%     \caption{Logic circuit of the gated SR latch. The SR latch within this circuit is represented by its symbol.}
%     \label{Gated SR logic scheme}
% \end{figure}
As previously observed, the SR latch is always transparent. This circuit can be further modified with an additional input, called \emph{latch enable} (LE), to develop a latch which becomes transparent only for a specific level of such input. This type of latch is often called a \emph{gated SR latch}. It represents an example of synchronous circuit and specifically defined as level-sensitive because its transparency depends on the level of the clock signal $LE$. The logic scheme of the gated SR latch is depicted in Fig. \ref{result GATED} (e-h). It is composed of an SR latch in which the $S$ and $R$ inputs pass through a first layer of NAND gates. The NAND gates take as inputs the actual set and reset of the gated latch, which we call $S'$ and $R'$, and the $LE$ input, which is in common between both gates. The overall behaviour of such device is shown in Truth Table \ref{result GATED}. A neuronal gated SR latch can be developed connecting the neuronal building blocks which implement the NAND gates and the SR latch following the same scheme of the gated SR latch, as illustrated in Fig. \ref{result GATED}.
% \begin{figure}[t]
% %\begin{center}
% \begin{tabularx}{\textwidth}{*{2}{>{\centering\arraybackslash}X}}
%    \centering
% \includegraphics[width=0.3\linewidth,valign=m]{Gated SR symbol.pdf} % <-- valing is chanded from T to m
% &
%      \begin{tabular}{c c c | c c}
%         $LE$ & $S'$ & $R'$ & $Q$ & $\overline{Q}$ \\
%         \hline
%         0 &  &  & \multicolumn{2}{l}{Memory state} \\
%         1 & 0 & 0 & \multicolumn{2}{l}{Memory state} \\
%         1 & 0 & 1 & 0 & 1 \\
%         1 & 1 & 0 & 1 & 0 \\
%         1 & 1 & 1 & \multicolumn{2}{l}{Not allowed} 
%         \end{tabular}         \\ % <-- added new row for captions
% \captionof{figure}{Symbol of the gated SR latch.}
% &     
%     \captionof{table}{\label{table gated SR}Truth Table of the gated SR latch.}
% \end{tabularx}
% % \end{center}
% \vspace{-10mm}
%  \end{figure}
 
% \begin{figure}[p!]
%     \centering
%     \includegraphics[width=0.65\textwidth]{Gated SR.pdf}
%     \caption{Functional scheme of the neuronal gated SR latch.}
%     \label{Gated SR scheme}
% \end{figure}

%SECTION
\subsection*{Neuronal flip-flops}

% \begin{figure} [b!]
%     \centering
%     \includegraphics[width=0.6\textwidth]{master_slave logic.pdf}
%     \caption{Logic circuit of the master-slave flip-flop.}
%     \label{Master_slave logic scheme}
% \end{figure}

% \begin{figure} [b!]
%     \centering
%     \includegraphics[width=0.6\textwidth]{D flip_flop logic.pdf}
%     \caption{Logic circuit of the D flip-flop.}
%     \label{D flip_flop logic scheme}
% \end{figure}

%Contrary to gated SR latch that are level-sensitive, i.e adopting the clock signal ($LE$) as a periodic square wave, the latch is transparent only during the high semi-periods of the clock \cite{pedroni2008digital}, flip-flop circuits are defined as edge-sensitive. They become sensitive only for a brief time, which is triggered by the transition of the clock between the two levels. In other words, the circuit samples the input values at the clock edge. Depending on the type of flip-flop, this could be the rising edge (\emph{positive-edge flip-flop}), the falling edge (\emph{negative-edge flip-flop}) or both of them (\emph{dual-edge flip-flop}).  

A straightforward modification of the primary-secondary flip-flop can be done in order to avoid the not-allowable condition. The $S'$ and $R'$ signals are substituted by a unique input signal $D$, which is sent directly for the $S$ input and inverted for the $R$ input, as depicted in Fig. \ref{result D FLIPFLOP7}. The above-mentioned flip-flop is called \emph{D flip-flop}, where D stands for data, because it requires a unique data input, or for delay, because the variations of the input are reported in the output after a certain delay defined by the next clock edge. The network structure that implements the neuronal D flip-flop is illustrated in Fig. \ref{result D FLIPFLOP7}. With this large circuit, the use of neuronal buffers becomes fundamental, especially in the neuronal NOT gates, due to the additional connection with the continuously firing neurons. Notice that all the sequential circuits presented in the current and previously in the results can be implemented by using AND NOT gates, NOT gates, AND gates, and neuronal buffers. Therefore, all the values of synaptic weights can be chosen according to the ones defined in each building block. On the whole, three values of synaptic weights are needed, especially $w_x$ and $w_y$, which are in common between the AND NOT gates, the NOTs and the buffers, and $w_z$, which is used for the synapses of the AND gates.

\subsection*{Neuronal Energy Model}

We use a modified eLIF model integrated with the Izhkevich neuron proposed in \cite{fardet2020simple}. The energy variable in the eLIF model is a proxy for the ATP/ADP ratio, reflecting the neuron's metabolic state. This variable affects the neuron's firing capability, with a lower energy state limiting the neuron's ability to emit spikes. %Energy dynamics are determined by a balance between energy consumption due to spike generation and maintenance of membrane potential, and energy production, which is primarily governed by mitochondrial oxidative phosphorylation.
The eLIF model thus provides a biophysically plausible method to study the interplay between neuronal electrophysiology and energy metabolism. It offers a valuable tool for exploring how energy limitations may contribute to the biocomputing burden posed onto the neurons. Considering

\begin{equation} \label{eq:energy}
\text{if } V < V_{th} \text{ or } \varepsilon < \varepsilon_{c}
\begin{cases}
C_m \dot{V} = g_L(E_L - V) + I_{syn} + I_{e} &  \\
\tau_{\varepsilon} \dot{\varepsilon} = (1 - \frac{\varepsilon}{\alpha \varepsilon_0})^3\frac{V - E_f}{E_d - E_f} & \text{else}\\
E_L = E_0 + (E_u - E_0)(1 - \frac{\varepsilon}{\varepsilon_0}) &  \begin{cases}
V \leftarrow V_r \\
\varepsilon \leftarrow \varepsilon - \delta
\end{cases}
\end{cases}
\end{equation}

\noindent where the membrane potential (\( V \)) is the electrical potential difference across the neuron's membrane, essential for the transmission of the neuronal signal. The membrane capacitance (\( C_m \)) reflects the neuron's ability to store and maintain charge, while the leak conductance (\( g_L \)) represents the propensity of the neuron to lose ions across its membrane, affecting the membrane potential. The leak reversal potential (\( E_L \)) is the equilibrium potential for the leak current. Synaptic currents (\( I_{syn} \)) represent the inputs from other neurons, and external input currents (\( I_e \)) corresponds to stimuli from outside the neuronal network. The energy variable (\( \varepsilon \)) is analogous to the ATP/ADP ratio and impacts the neuron's firing capability. The threshold potential (\( V_{th} \)) is the critical value the membrane potential must exceed for the neuron to fire an action potential, and the reset potential (\( V_r \)) is where the membrane potential is set post-spike. The time constant for the energy variable (\( \tau_{\varepsilon} \)) dictates the pace at which the energy state changes. Furthermore, \( \varepsilon_c \) indicates the critical energy level required for spike generation, while \( \alpha \) is a measure of the neuron's energetic health. The typical energy value (\( \varepsilon_0 \)) reflects the baseline energy state of the neuron, and \( \delta \) is the energy cost incurred by the neuron upon firing an action potential. 
\bibliographystyle{unsrt} 
% Bibliography
\bibliography{PNAS_BIOCOMP_ARXIV}

\end{document}